\documentclass[%
aps,
prl,
 amsmath,amssymb,nofootinbib,
 reprint,%
]{revtex4-1}
\usepackage{graphicx}
\usepackage{dcolumn}
\usepackage{bm}
\usepackage{color}

\def\beq{\begin{eqnarray}}
\def\eeq{\end{eqnarray}}

\begin{document}

\title{Can one hear the density of a drum? Weyl's law for inhomogeneous media} 
\author{Paolo Amore}\email{paolo.amore@gmail.com}
\affiliation{Facultad de Ciencias, CUICBAS, Universidad de Colima, \\
Bernal D\'{\i}az del Castillo 340, Colima, Colima, Mexico} 

\date{\today}

\begin{abstract}    
We generalize Weyl's law to inhomogeneous bodies in $d$ dimensions. 
Using a perturbation scheme recently obtained by us in Ref.~\cite{Amore09}, we have derived an explicit 
formula, which describes the asymptotic behavior of the eigenvalues of the negative laplacian on a closed 
$d$-dimensional cubic domain, either with Dirichlet or Neumann boundary conditions. For homogeneous bodies, the leading
term in our formula reduces to the standard expression for Weyl's law. 
We have also used Weyl's conjecture to obtain a non-perturbative extension of our formula and we have compared our analytical results
with the precise numerical results obtained using the Conformal Collocation Method of Refs.~\cite{Amore08,Amore09}.
\end{abstract}
\pacs{45.10.Db,04.25.-g} 
\maketitle

According to Weyl's law the largest frequencies of the sound of a uniform drum (membrane) 
are primarily determined by the area of the drum and not by its shape. 
The same result also applies to the electromagnetic field inside a waveguide, or to the 
eigenmodes of a particle in a quantum billiard, or more in general to the normal modes of
the Laplacian operator in a closed domain in $d$ dimensions, with Dirichlet or von Neumann 
boundary conditions.
 
Weyl conjectured that the number of modes in a two dimensional drum of area $A$ and perimeter $L$, 
of energy lower than a given energy $E$ is 
\beq
N(E) =  \frac{A}{4\pi} E \mp \frac{L}{4\pi} \sqrt{E} + \dots \nonumber
\eeq
where the $-$ and $+$ signs hold for Dirichlet and Neumann bc respectively. 
This formula is known as Weyl conjecture and, if true, it implies
\beq
E = \frac{4\pi N}{A} \pm \frac{L}{A} \sqrt{\frac{4\pi N}{A}} + \dots
\label{weylconjecture}
\eeq
which describes quite well the spectrum of a homogeneous drum. This conjecture has also been extended 
to $d$ dimension. Ref.~\cite{Steiner09}, which contains a detailed account 
of Weyl's law, provides a historical perspective. 

In this paper we want to ask ourselves a question similar to the one that Weyl posed himself:
how will an inhomogeneous drum, with density varying from point to point, sound? or, more in general,
what is the asymptotic behavior of the eigenvalues of the inhomogeneous Helmholtz equation
on a $d$-dimensional domain, with a spatially depending density, and with Dirichlet or Neumann
boundary conditions? And, finally, is it possible to justify Weyl's conjecture using perturbation theory?

To the best of our knowledge this problem has not been formulated before~\footnote{In Ref.~\cite{BH76} 
the inhomogeneous media problem is briefly mentioned in the section concerning "open problems".}, although its solution could find an 
incredible number of applications: for example to study of the acoustics of an inhomogeneous medium,
or the properties of an electromagnetic wave-guide with a varying dielectric coefficient, or even to study the propagation
of a seismic wave traveling in the interior of the earth. An asymptotic Weyl-type formula
for inhomogeneous bodies would also be helpful to extract informations on the density of the material composing the body,
thus allowing to rephrase Kac's question~\cite{Kac66} "can one hear the shape of a drum?" into "can one hear the density of a drum?"

Although these are difficult questions,  we will show in this paper 
that it is possible to  answer them, obtaining accurate estimates. We hope that the formalism devised here will
pave the way to more refined calculations and also provide an alternative way to look at the old and important problem
considered by Weyl.

Let us describe our approach. We focus for the moment being on two dimensional homogeneous membranes of arbitrary shape.
In a recent paper, Ref.~\cite{Amore09}, we have devised a perturbative approach to the calculation of the normal modes 
of these membranes  (or quantum billiards), that involves  a conformal map which sends the original shape into a reference 
shape (square, circle, etc.).

To be more explicit, $w=f(z)$ is a conformal transformation that maps a region $\Omega$, which for example could be chosen to be
a circle or a rectangle, into the region $\mathcal{D}$, representing the shape of the drum.
Under such transformation the homogeneous Helmholtz equation transforms to an inhomogeneous Helmholtz equation:
\beq
\Delta \psi(x,y) + E \Sigma(x,y) \psi(x,y) = 0 \ ,
\label{A5}
\eeq
where 
\beq
\Sigma \equiv \left| \frac{df}{dz} \right|^2 \ .
\label{Sigma}
\eeq
We will refer to $\Sigma$ as the {\sl conformal density}. Notice that $\Sigma$ could also be interpreted from the start as a
physical density, although the reverse is not necessarily true, as we shall soon see.

If we have in mind shapes which are obtained by small deformations of the reference shape $\Omega$, we may express
$\Sigma = 1 + \sigma$, where $\sigma$ is the {\sl perturbation density} generated by the mapping. In Ref.~\cite{Amore09}
we have obtained an explicit form for the perturbative corrections to the energy up to  third order in $\sigma$:
\beq
\label{pt11}E_n^{(0)} &=& \epsilon_n \\
\label{pt12}E_n^{(1)} &=& - \epsilon_n \langle n | \sigma | n \rangle  \\
\label{pt13}E_n^{(2)} &=& \epsilon_n \langle n | \sigma | n \rangle^2 
+ \epsilon_n^2 \sum_{k \neq n} \frac{\langle n | \sigma | k \rangle^2 }{\omega_{nk}} \\
E_n^{(3)} &=& - \epsilon_n \langle n | \sigma | n \rangle^3 + \epsilon_n^3 \langle n | \sigma | n \rangle
\sum_{k \neq n}  \frac{\langle n | \sigma | k \rangle^2}{\omega_{nk}^2} \nonumber \\
&-& 3 \epsilon_n^2  \langle n | \sigma | n \rangle \sum_{k \neq n}  
\frac{\langle n | \sigma | k \rangle^2}{\omega_{nk}} \nonumber \\
\label{pt14}&-& \epsilon_n^3 \sum_{k\neq n} \sum_{m \neq n} 
\frac{\langle n | \sigma | k \rangle \langle k | \sigma | m \rangle 
\langle m | \sigma | n \rangle }{\omega_{nk}\omega_{nm}}  \ .
\eeq
Clearly $\epsilon_n$ and $|n\rangle$ are the exact eigenvalues and eigenstates on $\Omega$ ($n$ is the set of
quantum numbers which define the state). We have defined $\omega_{nk} \equiv \epsilon_n-\epsilon_k$.

As we have observed in Ref.~\cite{Amore09} the first terms appearing in each of the equations above correspond
to the terms of a geometric series:
\beq
\epsilon_n \ \left( 1 - \langle n | \sigma | n \rangle + \langle n | \sigma | n \rangle^2  - \langle n | \sigma | n \rangle^3 + \dots\right) \ ,
\nonumber 
\eeq
which can be resummed as
\beq
E_n \approx \frac{\epsilon_n}{1 +   \langle n | \sigma | n \rangle} = \frac{\epsilon_n}{\langle n | \Sigma | n \rangle}  \ .
\label{weyl0}
\eeq

It is important to realize that, although the interpretation of $\Sigma$ as a conformal density is limited to two dimensional problems, 
the perturbative scheme of Ref.~\cite{Amore09} is general and it can be applied to a problem in $d$ dimensions, where $\Sigma$
is a {\sl physical} density. In what follows we will therefore work in $d$ dimensions, assuming $\Omega_d$ to be a $d$-dimensional
cube of side $2L$ centered in the origin and $\Sigma$ a physical density  inside $\Omega_d$.

The eigenfunctions of $\Omega_d$ corresponding to Dirichlet and von Neumann boundary conditions are obtained with the direct
product of the functions on each orthogonal direction. We have
\beq
\Psi^{(D)}_{n_x}(x) &=& \frac{1}{\sqrt{L}} \ \sin \left[\frac{n_x\pi}{2 L}(x+L)\right] 
\eeq
for the Dirichlet modes ($n_{x} = 1,2, \dots$), and
\beq
\Psi^{(N)}_{n_x}(x) &=& \frac{1}{\sqrt{L}} \  \cos \left[\frac{n_x\pi x}{2 L} - \frac{\pi}{4} (1-(-1)^n) \right] 
\eeq
for the von Neumann modes (this expression holds for $n_x=1,2,\dots$; for $n_{x}=0$, $\Psi^{(N)}_0(x)= 1/\sqrt{2L}$).

Let us now go back to eq.~(\ref{weyl0}): we first concentrate on the matrix elements $\langle n | \Sigma | n \rangle$,
and look for a suitable approximation for $n \rightarrow \infty$. As we have already mentioned here $n$ stands for the 
full set of quantum numbers specifying the states, i.e $n = (n_{x_1}, \dots, n_{x_d})$.

Therefore
\beq
\langle n | \Sigma | n \rangle = \int_{\Omega_d} {\Psi^{(D,N)}_{n_{x_1}}}^2(x_1) \dots {\Psi^{(D,N)}_{n_{x_d}}}^2(x_d)  \  \Sigma(x_1, \dots,x_d) 
\nonumber
\eeq

For $n_{x_i} \rightarrow \infty$ one may approximate the highly oscillatory functions with their average value, 
${\Psi^{(D,N)}_{n_{x_i}}}^2(x_i) \approx 1/2L$, and therefore write 
\beq
\langle n | \Sigma | n \rangle \approx \frac{1}{(2L)^d} \ \int_{\Omega_d} \Sigma(x_1, \dots,x_d) \ .
\nonumber
\eeq

We now would like to relate the energy of a state in $\Omega_d$, with quantum numbers $n_{x_1}, \dots, n_{x_d}$,
\beq
\epsilon_{n} \equiv \epsilon_{n_{x_1},n_{x_2}, \dots, n_{x_d}}  = \frac{\pi^2}{4L^2} (n_{x_1}^2+n_{x_2}^2+\dots + n_{x_d}^2)
\nonumber  \ ,
\eeq
to the number of states with  equal or lower energy $N(\epsilon)$.

As shown in Ref.~\cite{Dai09} the number of states with Dirichlet bc of energy less than  $E_{n_{x_1}, \dots, n_{x_d}}$ may be expressed as
\beq
N^{(D)}(E) &=& \frac{1}{(4\pi)^{d/2}} \left[ \frac{(2L)^d}{\Gamma(d/2+1)} \ E^{d/2} \right. \nonumber \\
&-& \left. \frac{(2 L)^{d-1} d \sqrt{\pi}}{\Gamma(d/2+1/2)} E^{d/2-1/2} +\dots \right] \ .
\eeq

For states obeying von Neumann bc we have
\beq
N^{(N)}(E) &\approx& N^{(D)}(E) \nonumber \\
&+& \frac{2}{(4\pi)^{d/2}} \frac{(2 L)^{d-1} d \sqrt{\pi}}{\Gamma(d/2+1/2)} E^{d/2-1/2}  \ ,
\eeq
where the second term counts the states with one of the quantum numbers vanishing.

Substituting these results in eq.~(\ref{weyl0}) we obtain a relation for the energy as a function of $N$:
\beq
E_N^{(D,N)} &\approx& \frac{\pi}{L^2} \frac{(2L)^d \left(\Gamma(d/2+1) \ N \right)^{2/d}}{\int_{\Omega_d} \Sigma(x_1,\dots, x_d)}  \nonumber \\
&\times& \left[ 1 \pm \frac{\left(\Gamma(\frac{d}{2}+1)\right)^{(d-1)/d}}{\Gamma(\frac{d+1}{2})} N^{-1/d} 
\right] + \dots
\label{weylsigma}
\eeq
which is a generalization of Weyl's law for a d-cube filled with density $\Sigma(x_1,\dots, x_d)$. The $+$ and $-$ signs hold for Dirichlet and Neumann bc respectively.

It is interesting to check some particular limits of this expression. For example, for $\Sigma=1$, one should recover the standard form of Weyl's law 
for an homogeneous d-cube: in this case the first term of eq.~(\ref{weylsigma}) reduces to  the correct expression
\beq
E_N^{(D,N)} \approx 4\pi \left(\frac{\Gamma(d/2+1) \ N}{V_d} \right)^{2/d} \ ,
\eeq
where $V_d = (2L)^d$ is the volume of the d-cube.

Let us now focus on the case $d=2$. In this case we have that 
\beq
E_N^{(D,N)} \approx \frac{4 \pi N}{\int_\Omega \Sigma(x,y) dxdy } \pm \frac{8 \sqrt{\pi N}}{\int_\Omega \Sigma(x,y) dxdy } + \dots
\label{weyl2D}
\eeq
which should be compared with eq.~(\ref{weylconjecture}). Assuming that $\Sigma$ is now the conformal density previously discussed
we may state Weyl's conjecture, eq.~(\ref{weylconjecture}), in the form
\begin{widetext}
\beq
E_N^{(D,N)} \approx \frac{4\pi N}{\int_\Omega \Sigma(x,y) dxdy } \pm
\frac{\int_{\partial\Omega} \Sigma^{1/2}(x,y) ds }{\int_\Omega \Sigma(x,y) dxdy } 
\sqrt{\frac{4 \pi N}{\int_\Omega \Sigma(x,y) dxdy }} + \dots  \ ,
\label{weylconjecture2}
\eeq
\end{widetext}
after noticing that $\int_\Omega \Sigma(x,y) dxdy$ and that $\int_{\partial\Omega} \Sigma^{1/2} (x,y) ds$ are 
the area and perimeter of the membrane. The reader may observe that eq.~(\ref{weylconjecture2}) reduces to our eq.~(\ref{weyl2D}) in the "perturbative" regime, $|\sigma|\ll 1$. As a matter of fact, in this limit one has that $\int_{\partial\Omega} \Sigma^{1/2}(x,y) ds /\sqrt{\int_\Omega \Sigma(x,y) dxdy} \approx 4$ (the perimeter to square root of the area ratio for a square), from which eq.~(\ref{weyl2D}) follows.

We will now use both eqns.~(\ref{weylconjecture}) and (\ref{weyl2D}) to study two dimensional drums of arbitrary shape and density.
The transformed Helmholtz equation now reads
\beq
\Delta \psi(x,y) + E \Sigma(x,y) \rho(u,v) \ \psi(x,y) = 0 \nonumber \ ,
\label{helmholtz}
\eeq
where $\rho(u,v)$ is the {\sl physical density} of the drum and $(u,v) = (Re f(z) , Im f(z))$. Therefore eqns.~(\ref{weylconjecture}) and (\ref{weyl2D})
must now be used substituting $\Sigma(x,y)$ with $\bar{\Sigma}(x,y) \equiv \Sigma(x,y) \rho(u,v)$. 

\begin{figure}
\begin{center}
\bigskip\bigskip\bigskip
\includegraphics[width=7cm]{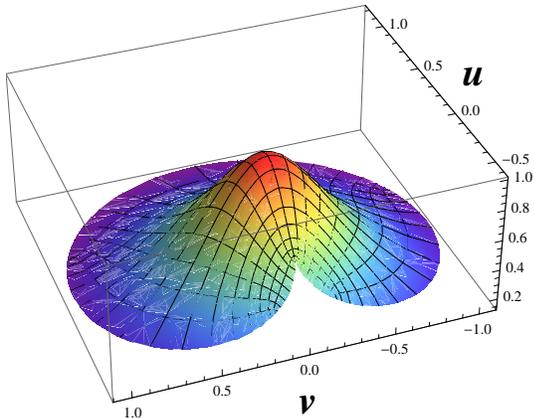}
\caption{(color online) Cardioid drum with density $\rho(u,v) = 1/(1+4 (u^2+v^2))$.}
\label{Fig_1}
\end{center}
\end{figure}

\begin{figure}
\begin{center}
\bigskip\bigskip\bigskip
\includegraphics[width=7cm]{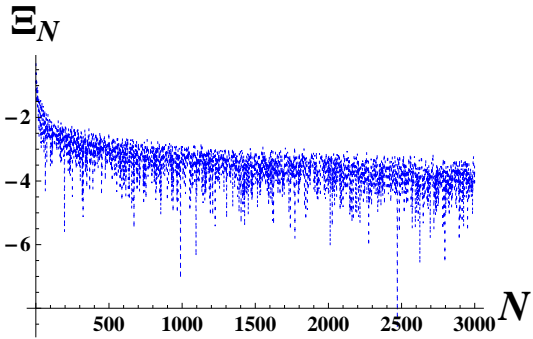}
\caption{(color online) The quantity 
$\Xi_{N} \equiv \log_{10} \left| 1- \left[\int_\Omega \bar{\Sigma}(x,y) dxdy\right] (E_N^{(D)}+E_{N}^{(N)})/8\pi N \right|$ 
for the cardioid drum of Fig.~\ref{Fig_1}. The energies $E_N^{(D)}$ and $E_N^{(N)}$ have been calculated using the 
Conformal Collocation method of Ref.~\cite{Amore08} with a grid of 99 points on each axis.}
\label{Fig_2}
\end{center}
\end{figure}

\begin{figure}
\begin{center}
\bigskip\bigskip\bigskip
\includegraphics[width=7cm]{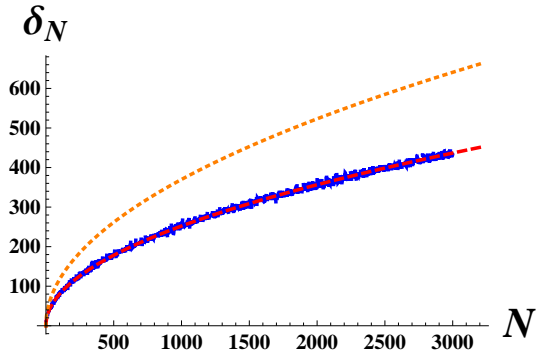}
\caption{(color online) The quantity $\delta_{N} \equiv (E_N^{(D)}-E_{N}^{(N)})/2$ for the cardioid drum of Fig.~\ref{Fig_1}.
The fluctuating (blue) line is the numerical result obtained using the Conformal Collocation method of Ref.~\cite{Amore08} with 
a grid of 99 points on each axis; the dashed (red) line is the second term in eq.~(\ref{weylconjecture2}) while the dotted (orange) 
line is the result obtained eq.~(\ref{weylsigma}).}
\label{Fig_3}
\end{center}
\end{figure}

We have tested eq.~(\ref{weyl2D}) and (\ref{weylconjecture2}) on a cardioid drum 
with density $\rho(u,v) = 1/(1+4 (u^2+v^2))$, as seen in Fig.\ref{Fig_1}. 
In this case we have that
\beq
\bar{L} &\equiv& \int_{\partial\Omega} \bar{\Sigma}^{1/2}(x,y) \ ds \approx 3.00112 \\
\bar{A} &\equiv& \int_\Omega \bar{\Sigma}(x,y) \ dxdy \approx 1.21205 \ .
\eeq

Notice that $\bar{L}$ and $\bar{A}$ do not have a geometric interpretation of perimeter and area
of a drum since the ratio $\bar{L}/\bar{A} = 2.48$ is smaller than the corresponding ratio 
between circumference and area of a circle with area $\bar{A}$, $2/\sqrt{\pi \bar{A}} = 3.22$.
Therefore $\bar{\Sigma}(x,y)$ cannot be obtained from a conformal transformation.

An independent confirmation of this observation comes from the Payne-Polya-Weinberger 
conjecture~\cite{PPW1,PPW2} (later proved by  Ashbaugh and Benguria in Ref.~\cite{Benguria91}), 
according to which the ratio between the first two eigenvalues of the Dirichlet laplacian is 
maximal for the circle, i.e.
\beq
\frac{E^{(D)}_2}{E^{(D)}_1} \leq \left.\frac{E_2}{E_1} \right|_{disk} = 
\left(\frac{j_{1,1}}{j_{0,1}}\right)^2 \approx 2.539
\label{C4}
\eeq
where $j_{0,1}$ and $j_{1,1}$ are the first positive zeroes of the Bessel functions $J_0(x)$ and $J_1(x)$.

Using the Conformal Collocation Method (CCM) with a grid of $99$ points in each direction we have found
for the inhomogeneous Helmholtz equation with density  $\rho(u,v) = 1/(1+4 (u^2+v^2))$ the values
\beq
E_1^{(D)} = 10.6769 \  \ , \ \ E_2^{(D)} = 29.7008 \ ,
\eeq
corresponding to a ratio $r = E_2^{(D)}/E_1^{(D)} \approx 2.78$. Since this ratio violates the Theorem 
proved by Ashbaugh and Benguria, $\bar{\Sigma}$ cannot be interpreted as a conformal density.
In other words, it is not possible to build an homogeneous drum of appropriate shape so that it sounds 
precisely as this inhomogeneous drum.

As we see in Fig.\ref{Fig_2} and \ref{Fig_3}, eq.~(\ref{weylconjecture2}) describes quite precisely the
behavior of the energies of this inhomogeneous drum. In particular from Fig.~\ref{Fig_2} we learn that
$(E_N^{(D)} + E_N^{(N)})/2$ essentially behaves as the first term in eq.~(\ref{weylconjecture2}), i.e.
{\sl Weyl's law for inhomogeneous drums}.

For bodies of dimension $d \geq 3$, our formulas only apply at present to cubical shapes of arbitrary density,
since it is not clear to the author if a generalization of a conformal transformation to higher dimensions exists. 
However, if this extension exists, there is an angle-preserving map which relates an arbitrary d-dimensional region to 
a d-cube. In this case the formula obtained in the present paper straightforwardly applies.

The results obtained here also allow to give a sense to the question contained in the
title: after the discovery made by Gordon, Webb and Wolpert in 1992, Ref.~\cite{Gordon92}, of two different 
homogeneous drums which are isospectral, it is now known that the answer to Kac's question is negative. 
In the present case, it make sense to wonder if there are truly genuinely inhomogeneous drums (i.e. inhomogeneous 
drums which cannot be reduced to a homogeneous one by means of a conformal transformation), which are isospectral
although corresponding to different (inequivalent) domains and/or densities. 

We conclude by observing that an improvement of the analytical formulas contained in this paper necessarily involves 
taking into account the terms in the perturbation expansion which have been neglected here. As these terms involve
sums over internal states, we expect them to be more difficult to evaluate; moreover, in this case the possible 
degeneration of the levels must also be taken into account, as discussed in Ref.~\cite{Amore09}. Most likely the calculation
of the corrections to the asymptotic formula of the present paper requires some averaging procedure over the population
of highly excited states. We plan to consider this issue in future works.

~\bigskip

\begin{acknowledgments}
It is a pleasure to thank Prof. A. Aranda for reading the manuscript.
I acknowledge support of Conacyt throught the SNI fellowship.
\end{acknowledgments}

\end{document}